\documentclass[times,tighten]{aastex631}
\shorttitle{EUV LATE PHASE IN HOMOLOGOUS FLARES}
\shortauthors{Zhong et al.}

\begin{document}
\title{Extreme-ultraviolet Late Phase in Homologous Solar Flares from a Complex Active Region}
\author[0000-0002-1203-094X]{Y.~Zhong}
\affiliation{School of Astronomy and Space Science, Nanjing University, Nanjing 210023, People's Republic of China}
\author[0000-0001-9856-2770]{Y.~Dai}
\affiliation{School of Astronomy and Space Science, Nanjing University, Nanjing 210023, People's Republic of China}
\affiliation{Key Laboratory of Modern Astronomy and Astrophysics (Nanjing University), Ministry of Education, Nanjing 210023, People's Republic of China}
\author[0000-0002-4978-4972]{M.~D.~Ding}
\affiliation{School of Astronomy and Space Science, Nanjing University, Nanjing 210023, People's Republic of China}
\affiliation{Key Laboratory of Modern Astronomy and Astrophysics (Nanjing University), Ministry of Education, Nanjing 210023, People's Republic of China}

\correspondingauthor{Y.~Dai}
\email{ydai@nju.edu.cn}

\begin{abstract}
Recent observations in extreme-ultraviolet (EUV) wavelengths reveal a new late phase in some solar flares, which is seen as a second peak in warm coronal emissions ($\sim3$ MK) several tens of minutes to a few hours after the soft X-ray (SXR) peak. The origin of the EUV late phase (ELP) is explained by either a long-lasting cooling process in the long ELP loops, or a delayed energy ejection into the ELP loops well after the main flare heating. Using the observations with the \emph{Solar Dynamics Observatory} (\emph{SDO}), we investigate the production of the ELP in six homologous flares (F1--F6) originating from a complex active region (AR) NOAA 11283, with an emphasis on the emission characteristics of the flares. It is found that the main production mechanism of the ELP changes from additional heating in flare F1 to long-lasting cooling in flares F3--F6, with both mechanisms playing a role in flare F2. The transition is evidenced by an abrupt decrease of the time lag of the ELP peak, and the long-lasting cooling process in the majority of the flares is validated by a positive correlation between the flare ribbon fluence and the ELP peak intensity. We attribute the change in ELP production mechanism to an enhancement of the envelope magnetic field above the AR, which facilitates a more prompt and energetic heating of the ELP loops. In addition, the last and the only confined flare F6 exhibits an extremely large ELP. The different emission pattern revealed in this flare may reflect a different energy partitioning inside the ELP loops, which is due to a different  magnetic reconnection process.
\end{abstract}

\keywords{Solar flares (1496); Solar extreme ultraviolet emission (1493); Solar active regions (1974)}

\section{Introduction}
Solar flares, manifested as transient enhancements of electromagnetic radiation over a wide wavelength range, are one of the most energetic phenomena in the solar atmosphere. The energy capable of driving a solar flare comes from the coronal magnetic fields. Through the magnetic reconnection process \citep{Parker63}, the non-potential magnetic energy stored in the corona is rapidly converted into plasma heating, particle acceleration, and in case of an eruptive event, large-scale mass motion that is usually known as a coronal mass ejection \citep[CME, e.g.,][]{Priest02,Fletcher11}.

According to the standard two-ribbon solar flare model \citep[][conventionally termed the CSHKP model]{Carmichael64,Sturrock66,Hirayama74,Kopp76}, energy release during the ongoing magnetic reconnection instantly heats up the newly reconnected post-flare loops (flare loops for short), whose evolution can be divided into two phases as typically seen in soft X-ray (SXR) emissions: an impulsive rise, followed by a more gradual decay toward the background level. The two phases reflect a dynamic heating-cooling cycle of the flare loops. The impulsive phase coincides with the impulsive heating and a follow-up evaporation of hot material driven by thermal conduction and/or non-thermal electron bombardment \citep{Antiochos78}, and the gradual phase corresponds to a radiative-dominated cooling together with enthalpy draining during the later stage \citep{Antiochos80,Bradshaw10}.

Emission from the extreme-ultraviolet (EUV) domain provides another important tool to diagnose the flare evolution. In the CSHKP model, the EUV coronal emissions of a solar flare resemble the tendency seen in SXR, with the emission peaks in different lines/passbands occurring sequentially in an order of decreasing temperatures, coincident with the cooling of the flaring plasma \citep{Chamberlin12}. Nevertheless, by using full-disk integrated EUV irradiance observations with the EUV Variability Experiment \citep[EVE;][]{Woods12} on board the recently launched \emph{Solar Dynamics Observatory} \citep[\emph{SDO};][]{Pesnell12} mission, \citet{Woods11} discovered a new ``EUV late phase (ELP)" in some solar flares, which is predominately seen as a second peak in the warm coronal emissions (e.g., \ion{Fe}{16}~335 {\AA}, $\sim$3 MK) several tens of minutes to a few hours after the \emph{GOES} SXR peak. Furthermore, spatially resolved observations as taken with the Atmospheric Imaging Assembly \citep[AIA;][]{Lemen12} also on board \emph{SDO} reveal that the ELP emission originates from a set of higher and longer flare loops rather than the main flare loops, even though the two groups of loops are magnetically related within the same active region (AR).

In a preliminary statistical study, \citet{Woods11} found that ELP flares are not so common among all solar flares, and they tend to occur from some specific ARs. This has been later validated by many case studies that reveal a multipolar magnetic configuration of the flare-hosting ARs: they exhibit either a quadrupolar configuration \citep{Hock12,LiuK13,WangY20}, or a parasitic polarity embedded in a large-scale bipolar magnetic field \citep{Dai13,Dai18b,SunX13,Masson17}. In particular, the latter configuration facilitates the formation of a circular-shaped flare ribbon together with a remote ribbon, which implies a three-dimensional (3D) fan-spine topology as confirmed by force-free field extrapolations \citep{SunX13,Jiang14,LiYD14,Masson17}, with the field lines in distinct lengths tracing different sets of flare loops.

Upon discovery, the ELP was at first attributed to an additional heating in the ELP loops that takes place well after the heating of main flare loops \citep{Woods11,Hock12,Dai13,Zhou19}. This delayed heating should be much weaker compared to the main flare heating, therefore mainly enhancing the emissions at intermediate temperatures. Based on the fact that the ELP loops are significantly longer than the main flare loops, an alternative scenario was proposed that both sets of flare loops are heated simultaneously during the main phase, but the ELP loops experience a much more extended cooling process \citep{LiuK13,Masson17,Dai18b}. The difference in cooling time scale between the loops of different lengths can naturally explain the separation of an ELP peak from the main phase peak \citep{Cargill95,Chen20}.  It was also pointed out that the two mechanisms may both play a role in an ELP flare \citep{SunX13}.

Using the zero-dimensional (0D) hydrodynamic model enthalpy-based thermal evolution of loops \citep[EBTEL;][]{Klimchuk08,Cargill12,Barnes16}, \citet{Dai18a} have recently numerically probed the production of ELP flares. It was found that under the above two mechanisms, the ELP loops may experience different cooling stages around the ELP peak: radiative cooling stage for the long-lasting cooling scenario while conductive cooling stage for the additional heating scenario. This difference in loop thermodynamic evolution will cause opposite patterns in the shape of the synthetic flare light curves, which can be in turn used as a new diagnostic tool to differentiate the two mechanisms from real solar observations.

The origin of ELP in solar flares is still not fully understood. Since the occurrence of ELP flares shows a tendency to be in clusters from the same AR, a comparative analysis of homologous ELP flares under the similar magnetic environment will certainly shed new light on this issue. In this paper, we focus on such flares from a complex AR, with an emphasis on the emission characteristics of the ELP in these flares. The rest of the paper is organized as follows. In Section 2 we give a brief introduction to the instruments and data used in this study. Detailed analysis of the observational data is carried out in Section 3. The results are discussed and main conclusions are drawn in Section 4.

\section{Instruments and Dataset}
The data used in this work are primarily from the instruments on board \emph{SDO}. EVE measures full-disk integrated solar irradiance over a wavelength range between 1 {\AA} and 1050 {\AA} with 1 {\AA} spectral resolution  and 10 s cadence. As a subset of EVE data products, the  irradiance of some ``isolated" lines is derived by integrating the EVE spectra over specified spectral windows. AIA provides simultaneous imaging observations of the solar corona and transition region (TR) in 10 EUV/UV passbands with a pixel size of 0.6$\arcsec$ and a temporal resolution of 12 s or 24 s. In addition, magnetograms obtained with the Helioseismic and Magnetic Imager \citep[HMI;][]{Scherrer12} help us check the magnetic configuration of the flare-hosting AR.

We focus on a complex AR NOAA 11283 that was recorded passing through the visible disk from 2011 August 31 to September 11. The AR harbored a series of major flares, which have been extensively studied in the literature \citep[e.g.,][]{Feng13,Jiang14,LiuC14,Xu14,Romano15,Ruan15,ZhangQM15}. In particular, some of the flares also exhibited a notable ELP \citep[see][]{Dai13,Dai18b}.

For event selection, we apply two following  criteria. First, we choose flare events above the M class to ensure that the flares are energetic enough to power a potential ELP.  As reported in previous studies, an ELP can be observed in flares as weak as the C2 level \citep{Woods11}. However, after carefully checking all flares in AR 11283, we did find that none of the flares below our selection threshold produced a detectable ELP. Second, the flares should be located within a heliocentric distance of $60^{\circ}$ from the disk center so that the identification of flaring structures is less affected by the effects of foreshortening and/or line-of-sight (LOS) overlapping.  As revealed by the \emph{GOES} 1--8~{\AA} SXR time profile shown in Figure 1, there were a total of 7 M/X-class flares (labeled as F1 to F7) originating from AR 11283. Among these 7 candidate events, the last one (F7 on September 10) was quite close to the western limb (N12W61), hence being removed from our final dataset.

Some general information on the selected flares and their associated phenomena are listed in Table 1. As seen in the table, and also revealed in Figure 1, the magnitude of the major flares from AR 11283, with the exception of flare F1, shows a tendency of decrease with time, which may reflect a general decay of the AR energetics with the flare energy release.  With the aid of other data sources such as those from the Large Angle and Spectrometric Coronagraph \citep[LASCO;][]{Brueckner95} on board the \emph{Solar and Heliospheric Observatory} (\emph{SOHO}) , it is found that all flares except for the last one (flare F6) are associated with a CME, implying eruptive events. Among the five eruptive flares, all but one (flare F4)  are accompanied by a metric type II radio burst, meaning that most of the CMEs are fast enough to drive a coronal shock.

\section{Observations and Results}
\subsection{Overview of the AR and Flares}
Figures 2(a)--(c) and (g)--(i) show the HMI LOS magnetograms of AR 11283. Upon appearance, the AR was largely a bipolar one, with two main polarities labeled as P1 (for positive) and N (for negative), respectively. Under such simple magnetic configuration, the flare level of the AR was quite low during the early stage. From September 4 to 5, a persistent emergence of opposite magnetic flux took place inside the leading polarity N, finally evolving into a new parasitic polarity P2 surrounded by polarity N. This flux emergence significantly complicated the magnetic configuration of the AR, leading to the occurrence of the consequent major flares studied in this work.

As seen from the AIA 1600 {\AA} chromospheric images in Figures 2(d)--(f) and (j)--(l), under the similar multipolar magnetic configuration, the flares also exhibit similar multiple ribbon morphology. The flare ribbons are predominately manifested as a circular ribbon in the western negative polarity N (labeled as Rc) and an elongated ribbon in the eastern positive polarity P1 (labeled as Rr). The two flare ribbons brighten up and reach the peak nearly simultaneously (typically within 1 minute, see Table 2). At first glance, they seem to constitute a pair of conjugate ribbons as typically seen in a two-ribbon flare. Nevertheless, a third rather compact ribbon can be identified in the parasitic polarity P2, which is much closer to and roughly parallel with the circular ribbon Rc. Affected by saturation of the instrument detectors in strongly flaring regions, this ribbon is less distinguishable from ribbon Rc. As we will see below, all the flares are driven by the eruption of a sigmoid filament initially lying along the polarity inversion line (PIL) between polarities P2 and N (hereafter PIL-P2N). In this sense, these two adjacent ribbons should belong to main flare ribbons, while the eastern flare ribbon Rr is more likely to correspond to a remote ribbon. Because of the general similarity between the flares, we call them homologous flares.

Figure 3 displays the background-subtracted irradiance of the flares in three EVE lines, including the \ion{Fe}{20} 133 {\AA} ($\sim$10 MK), \ion{Fe}{16} 335 {\AA} ($\sim$3 MK), and \ion{Fe}{9} 171 {\AA}  ($\sim$0.8 MK), which cover emissions from the hot, warm, and cool corona, respectively. In each flare, the hot coronal \ion{Fe}{20} 133 {\AA} emission closely resembles the \emph{GOES} SXR light curve due to their similar temperature response. For the eruptive flares (F1--F5), the cool coronal \ion{Fe}{9} 171 {\AA} emission quickly turns into a dimming after the main phase peak, which may reflect a mass drainage during the CME lift-off \citep[e.g.,][]{Aschwanden09,ChengJ16}. Of particular interest is another enhancement of the warm coronal \ion{Fe}{16} 335 {\AA} emission well after the main phase peak, indicative of a potential ELP according to \citet{Woods11}. The emission pattern of the ELP seen in EVE 335 {\AA} is quite different in different flares. In flares F1 and F2, the ELP shows an extended decay. In flare F3, the ELP has a plateau-like shape followed by a relatively fast drop. In flares F4 and F5, the EVE 335 {\AA} irradiance after the main phase peak seems to fall back to an elevated ``background", from which the identification of a possible ELP needs to further resort to spatially resolved observations with AIA. For the only non-eruptive flare F6, it exhibits an extremely large ELP, with peak emission over 1.3 times that of the main phase peak.

\subsection{Spatially Resolved Observations of Individual Flares}
\subsubsection{The September 6 X2.1 Flare}
Rather than following the exact chronological order of flare onset, we start our analysis with flare F2 (September 06 X2.1), an event having been investigated  in our previous work \citep{Dai13}, whose general evolution has also been reproduced by \citet{Jiang18} using a data-driven magnetohydrodynamic (MHD) numerical simulation. Here we extend our previous study and revisit the origin of the ELP in this flare, which will serve as a basis for the analysis of other events.

The left panels of Figure 4 show characteristic snapshots of the flare evolution in the AIA passbands of 131, 335, and 171~{\AA}, respectively, the choice of which is based on a general temperature correspondence to the three EVE lines shown in Figure 3. The flare begins with the eruption of an east–west (EW) oriented filament from the southern part of PIL-P2N around 22:18 UT (top row in Figure 4). As pointed out by the arrow in Figure 4(e), the filament exhibits a forward ``\textsf{S}" shape, indicative of a magnetic flux rope (FR) of the sinistral chirality \citep{ChenPF14}, which is properly reconstructed lying under the dome of a fan-spine magnetic structure in non-linear force-free field (NLFFF) extrapolations  \citep{Feng13,Jiang14}. We note that this northern hemisphere filament does not obey the well-known hemisphere rule that the chirality of a filament is mainly dextral/sinistral in the northern/southern hemisphere \citep[e.g.,][]{Ouyang17}, thus being more plausibly prone to a major eruption (P.~F.~Chen, private communication).

Like in a typical two-ribbon flare, the fast rising FR stretches out the overlying field lines to drive standard CSHKP flare reconnection, producing two parallel flare ribbons and bright main flare loops along the southern part of PIL-P2N\@. In this case, however, the FR is initially embedded under a fan-spine structure. When it erupts upward, the FR also drives other magnetic reconnections at the magnetic null-point intersected by the fan-spine fields lines as well as in the quasi-separatrix layer (QSL) surrounding the fan dome. As a result, the outer parallel flare ribbon rapidly evolves into a complete circular ribbon (Rc), and a remote flare ribbon (Rr) appears in the eastern polarity P1, at a distance of 80$\arcsec$ from the main flare region. Meanwhile, the region enclosed by the remote flare ribbon quickly turns into a dark dimming, which is due to mass drainage along the large-scale field lines stretched out by the erupting FR \citep[e.g.,][]{Xing20}.

The second-stage flare evolution is characterized by the ejection of another filament lying along the northern part of PIL-P2N about 15 minutes later (second row in Figure 4), whose destabilization may be triggered by a removal of the above magnetic confinement by the first-stage FR eruption. The magnetic reconnection driven by this moderate filament ejection causes brightening of two sets of side-lobe flare loops (most clearly seen in AIA 131 {\AA}) south and east of the uprising filament, respectively, analogous to the breakout reconnection as proposed by \citet{Antiochos99}.

In the last stage, much longer flare loops connecting the large-scale bipolar polarities P1 and N are observed (bottom two rows in Figure 4). Since these loops appear in AIA 335 {\AA} well after the main phase peak (over an hour), they definitely belong to ELP loops. In AIA 335 {\AA}, the ELP loops are rather diffuse over a large area, with their eastern footpoint  rooted on the previously brightening remote flare ribbon Rr. Nevertheless, in the other two AIA passbands, the ELP loops are seen as relatively sharp structures. Their appearance in these passbands is first localized within a small region (limited by the black dashed rectangle), and later spreads to a wider extent (outlined by the white dashed rectangle). Note that during this late stage, the AIA 131 {\AA} passband should be mainly sensitive to cool rather than hot coronal emissions \citep[c.f.,][]{ODwyer10}.

The three stages in flare evolution identified in the spatially resolved AIA observations have a one-to-one temporal correspondence to the three enhancements of the EVE 335 {\AA} irradiance in Figure 3(b), validating the identification of the ELP in flare F2 from the EVE observations. Actually, the evolution of the AR intensity in each AIA passband, which is derived by summing the count rate of all pixels over the AR (plotted in Figure 4(m)), shows a close similarity to that of the full-disk irradiance in the corresponding EVE line, suggesting that the variabilities measured in EVE are predominately contributed by the activities taking place in the AR. We note that this general similarity holds for most of the flare events studied in this work.

To explore the origin of the ELP in flare F2, we select three sub-regions enclosed by the colored boxes drawn in Figures 4(f) and (g), representing  the main flare region (labeled as R1 in the red box), the ELP loop region (R2 in blue), and the remote dimming region (R3 in purple), respectively. The background-subtracted light curves of the three sub-regions in AIA 335 {\AA} are plotted in Figure 4(n). As expected, the activities in region R1 dominate the first two peaks in the AR-integrated (or whole flare) light curve, while the variabilities in region R2 make a main contribution to the ELP (the third enhancement). Note that the former two peaks in the R2 light curve, which are simultaneous with the peaks in the R1 light curve, are mainly caused by the contamination of CCD-bleed interference stripes extending from the main flare region.

Flare F2 exhibits an extended ELP seen in AIA 335 {\AA}, indicating a prolonged heating during this stage. As revealed from the online animation as well as Figure 3 in \citet{Dai13}, the additional heating during the ELP is evidenced by re-brightening and eastward expansion of the remote ribbon Rr in the AIA coronal passbands, a signature of ongoing magnetic reconnection. This magnetic reconnection takes place between the high-lying field lines stretched out by the preceding filament eruption/ejection, analogous to the main flare CSHKP reconnection but in a much weaker strength. In the AIA 1600~{\AA} light curves shown in Figure 4(o),  potential enhancements in response to the ELP reconnection are barely discernible. Nevertheless, powered by this weak but persistent reconnection heating, the background-subtracted intensity over region R3 gradually turns from a depletion shortly after the main phase peak to an enhancement, which is mainly contributed by the brightening of the ELP loops at their eastern leg. As the individual ELP loops cool down, they successively appear and disappear in the cooler passbands such as AIA 171 {\AA} (Figure 4(l)), with such process lasting until the first few hours of the next day, further validating the scenario of additional delayed heating.

In our previous study \citep{Dai13}, the production of the ELP in this flare was just attributed to the additional heating. Nevertheless, the brightening of the remote flare ribbon Rr in all three stages suggests that some ``ELP" loops anchored on ribbon Rr have already been heated early in the flare. Powered by the much stronger early stage reconnection heating, these long flare loops should be initially quite hot ($>$10 MK) as seen in AIA 131 {\AA} (Figures 4(a) and (b)). After experiencing a long cooling process, they are observed as conventionally defined ELP loops in AIA 335 {\AA} much later (Figure 4 (g)), and then appear or reappear in AIA 131  (now for cool corona, $<$1 MK) and 171 {\AA} (enclosed by the black dashed rectangle in Figures 4(c) and (k)). Alternatively, the production of these ELP loops should conform to the long-lasting cooling scenario.

To further verify the long-lasting cooling process in these ELP loops, we select such a representative loop from the region limited by the black dashed rectangle drawn in Figure 4(c), and trace its evolution in Figure 5. As shown in the left panels in Figure 5, the loop has already brightened up in AIA 131 {\AA} during the first stage of flare evolution (Figure 5(b)). After a sufficiently long cooling time ($\sim$87 minutes), the loop reappears in AIA 131 {\AA} in the last stage (Figure 5(c)). By visually checking the loop morphology frame by frame, we make sure that the loop structure seen at the two different moments belongs to the same ELP loop but with different temperatures. We draw a $1.2\arcsec\times1.2\arcsec$ sized small box on the loop spine, over which the intensity profiles in AIA 131 and 335 {\AA} are plotted in Figure 5(d). It is found that the observation times of Figures~5(b) and 5(c) correspond to two local peaks in the AIA 131 {\AA} light curve (outlined by the vertical dashed lines).  Note that other peaks in the AIA 131 {\AA} intensity profile should be due to the contribution from other ELP loops overlapping along the LOS\@.

It is also seen that the AIA 335 {\AA} intensity profile of the box region shows a slow rise followed by a fast decay during the period of interest. According to the numerical experiments carried out in \citet{Dai18a}, if an ELP loop is energized simultaneously with the main flare loops, it will be initially heated to a high temperature and hence experience a ``complete" heating-cooling cycle. As a result, the warm coronal emission peak (e.g., in AIA 335 {\AA}) in the ELP loop will occur in the radiative cooling stage of the loop, and the loop emission after the peak will be notably affected by mass condensation from the loop, hence showing a fast decay compared to the relatively gradual rise. This numerically predicted emission pattern is qualitatively consistent with the tendency revealed in the AIA 335 {\AA} light curve.

To quantify the cooling process of the loop, we further carry out a differential emission measure (DEM) analysis to the selected box region using the observations from six AIA coronal passbands. Here we adopt the sparse inversion code developed by \citet{Cheung15} and modified by \citet{Su18} to perform the DEM inversion. The resulting DEM distributions at the two AIA 131 {\AA} peak times (blue solid histogram) are plotted in Figures~5(e) and (f). Also plotted is the DEM at a pre-flare time (black solid histogram), which serves as a reference DEM for comparison. At the former (early) AIA 131 {\AA} peak (22:24~UT), the DEM distribution has a prominent bump between $\log T\sim6.8$ and $\log T\sim7.2$, outside which the DEM closely resembles the reference  pre-flare DEM (Figure 5(e)). Obviously this high temperature component originates from the ELP loop that has just been heated to a high temperature. Hence we can isolate this component out of the whole temperature range to estimate the parameters of the heated ELP loop. Adopting the DEM-weighted average, and assuming an LOS loop depth (width) of 1.5 Mm, the derived average loop temperature $\overline T$ and density $\overline n$ are 11~MK and $1.5\times10^{10}$~cm$^{-3}$, respectively, typical of a flaring coronal loop. At the latter (late) AIA 131 {\AA} peak (23:51~UT), nevertheless, the high temperature bump totally disappears from the DEM histogram, and enhancements of the DEM are mainly observed at lower temperatures ($\log T<6.3$, Figure 5(f)). The change in DEM distribution clearly demonstrates the long-lasting cooling (over one hour) of the selected ELP loop from a flaring plasma temperature to the background coronal temperature. In passing we note that there are several missing bins in the DEM plots at around $\log T=5.9$, in which the DEM values are orders of magnitude smaller than those in neighboring bins. This might be a problem with the inversion algorithm itself. Hence it does not physically mean that the DEM at these temperatures is low. Fortunately, this issue would not compromise the results of our DEM analysis, because we intend to extract quantities of the ELP loop out from the background corona, especially when it is of high temperature.

\citet{Cargill95} analytically evaluated the cooling time of a flare loop (cooling down from a high temperature), which is formulated as
 \begin{equation}
 \label{EQ}
\tau_{\mathrm{cool}}=2.35\times10^{-2}L^{5/6}T_0^{-1/6}n_0^{-1/6},
\end{equation}
where $L$ is the loop half-length, and $T_0$ ($n_0$) is the loop temperature (density) at the start of the cooling. We use this formula to estimate the cooling time of the selected ELP loop. By measuring the distance between two footpoints of the loop and assuming a semi-circular loop geometry, the loop half-length is estimated to be  $\sim7.2\times10^9$ cm. As to $T_0$ and $n_0$, we adopt the values derived from the DEM at the first AIA 131 {\AA} peak. Then the theoretically estimated loop cooling time is 87 minutes, which is almost the same as the observed cooling time (approximated by the duration between the two peaks in the AIA 131 {\AA} light curve for the box region), therefore serving as another piece of compelling evidence for the mechanism of long-lasting cooling.

In summary, it is found that both the additional heating and the long-lasting cooling mechanisms work in the production of the ELP in flare F2. The ELP loops produced through the two mechanisms overlap temporally and spatially around the ELP peak, making it difficult to quantify the relative contribution from each mechanism. We just speculate that at least at the ELP peak, both mechanisms may play a comparable role.

\subsubsection{The September 6 M5.3  Flare}
We then turn to flare F1 (September 6 M5.3), the first major flare event from AR 11283. As shown in Figure 6, the flare, involving a similar fan-spine topology, experiences nearly the same three-stage evolution as in flare F2, which is also characterized by sequential occurrence of the eruption of an FR (from the southern part of PIL-P2N, upper panels), the ejection of a filament (from the northern part of PIL-P2N, middle panels), and the brightening of ELP loops (mainly in region R2, lower panels). This implies that similar magnetic reconnections could be involved in this flare. However, in each stage, the involved reconnection should be significantly weaker than that in flare F2. This is reflected from the remote flare ribbon Rr in flare F1, which is considerably weak and compact, even during the strongest first-stage reconnection (see Figure 2(d)). As a result, possible ``hot" ELP loops rooted on ribbon Rr are barely discernible in AIA 131 {\AA} during the early stages (Figures 6(a) and (b)), hence ruling out the possibility of long-lasting cooling as the main mechanism to produce the warm coronal ELP later on.

In AIA 335 {\AA}, the brightening of ELP loops occurs much later, and persists for a long time (Figure 6(f)). Like in flare F2, the AIA 335 {\AA} intensity profiles for both regions R2 and R3 (defined the same as in flare F2) show an extended slow rise during the ELP (Figure 6(k)), strongly favoring the additional heating scenario in producing the extended ELP\@. It is noted that during the ELP period, the AR-integrated light curve in AIA 335 {\AA} mainly reveals a depletion (Figure 6(j)) rather than an enhancement seen in EVE 335 {\AA} (Figure 3(a)). This can be understood in terms of the broad temperature response of the AIA 335 {\AA} passband as well as the weak strength of the delayed ELP heating. Besides warm coronal emissions, AIA 335 {\AA} also covers emissions from the bulk cool corona \citep[c.f.,][]{Boerner12}. In flare F1, the filament eruption/ejection causes a significant mass drainage from the background/foreground AR, leading to a relatively deep dimming observed in AIA 335 {\AA}. Although the late-stage reconnection heating can make the ELP loops brighten up in a local region within the AR, the heating is so weak that the brightening from the ELP loops is not enough to compensate for the overall dimming.

\subsubsection{The September 7 X1.8 Flare}
According to the \emph{GOES} classification, flare F3  (September 7 X1.8) is the second most energetic flare in our event list. The evolution of the flare is displayed in Figure 7. Compared to the former two flares, flare F3 is  driven by the eruption of just one single sigmoid filament (or FR, indicated by the arrow in Figure 7(e)), which is also the case for the remaining flare events to be analyzed. For this flare, the NLFFF extrapolation also reveals a fan-spine magnetic structure lying above the initially EW oriented FR, with the exception that the connectivity of the outer spine experiences a change between flares F2 and F3: the outer spine still extends to the northwest shortly after flare F2 \citep[see][]{Jiang14}, while it has connected northeastward to the eastern positive polarity P1 before flare F3 (Y. Dai, in preparation).  Due to the similarity, the fast rising FR first drives similar magnetic reconnections as in flare F2, i.e., main flare CSHKP reconnection (below the FR) as well as null-point reconnection (above the FR). Evidence of the null-point reconnection is further revealed from the energization of a bundle of flare loops brightening up in AIA 335 {\AA} from $\sim$22:50 UT extending northeastward to the remote ribbon Rr (see Figure 7(i) and the online animation), which are roughly co-spatial with the outer spine filed lines in the NLFFF extrapolation. As the bundled flare loops cool down, they are more clearly observed in AIA 171 {\AA} (covered by the red box in Figure 7(j)). We note that the situation is very similar to the production of hot spine loops by null-point reconnection as proposed in \citet{SunX13}.

As the FR erupts upward, the FR spine further undergoes a clockwise rotation, whose direction can be determined from the locations of falling filament material onto the solar surface \citep{Ouyang15}. The sense of the rotation follows the chirality rule, i.e., counterclockwise for a dextral filament while clockwise for a sinistral filament \citep{Attrill07}. As a result of the rotation, the FR changes from the initial EW orientation to a later north-south (NS) orientation. The change in FR orientation imposes a significant consequence to the follow-up magnetic reconnection. Since the large-scale overlying field lines (above the fan-spine dome) are mainly EW directed, an NS elongated FR can more effectively stretch out a large amount of overlying field lines, facilitating stronger magnetic reconnection between them. Without further stretching by other filament eruption/ejection, magnetic reconnection (analogous to the main flare CSHKP reconnection) quickly sets in between the stretched overlying field lines, producing a large-scale cusp-shaped structure \citep[a signature of magnetic reconnection,][]{Tsuneta92,Aschwanden05} with one footpoint anchored on the remote ribbon Rr (Figure 7(b)). Although the main flare reconnection (creating the smaller ribbons along polarities P2 and N) and the large-scale CSHKP reconnection (producing the large-scale cusp straddling on the longer remote ribbon) are spatially separated, the two reconnections are closely linked by the erupting FR\@. The FR starts to erupt at $\sim$22:36~UT, while the large-scale hot cusp structure appears as early as 22:41 UT\@. This means that the transition in reconnection location happens within a short time interval (less than 5 minutes). In addition, driven by the FR eruption, both reconnections last for a period of time, hence showing an overlap in time. In this sense, the two reconnections can be regarded to take place simultaneously.

As revealed from both the full-disk irradiance in EVE 335 {\AA} (Figure 3(c)) and the AR-integrated intensity in AIA 335 {\AA} (Figure 7(m)), flare F3 exhibits a plateau-like ELP\@. Through visual inspection of the AIA imaging observations, it is seen that the plateau ELP emission comes from a series of ELP loops of different lengths, which are heated by the above-mentioned null-point reconnection or large-scale CSHKP reconnection. Since all reconnections in this flare take place nearly simultaneously, both the main flare loops and the ELP loops should also appear simultaneously in the hot AIA passbands, as is the case seen in AIA 131 {\AA} (Figures 7(a) and (b)). As the ELP loops cool down, the difference in loop length leads to a difference in cooling rate of the ELP loops, which makes the brightening of different ELP loops more and more separated in time, especially seen in the cool AIA 171 {\AA} passband (Figures 7(j)--(l)). In the warm AIA 335 {\AA} passband, nevertheless, the time lags between the emission peaks of ELP loops are not so large. Therefore, the ELP loops of different lengths brighten up and fade off in close succession (Figures 7(f)--(h)), making the overall warm coronal ELP emission rather flatted.

We further pick up three sets of ELP loops (I, II, and III) that have increasing lengths. Note that loops I correspond to the hot spine energized by the null-point reconnection, and loops II and III are related to the initially hot cusp produced by the large-scale CSHKP reconnection. In each set of loops, we extract a segment region (outlined by the colored boxes R1, R2, and R3) to trace the loop intensity in AIA 335 {\AA}. As displayed in Figure 7(n), the AIA 335 {\AA} intensity profiles of the three loop segments exhibit a gradual rise shortly after the main phase peak, attain the peak in an order of increasing loop lengths, and then turn to a relatively faster decay. The pattern of the light curves is quite similar to that plotted in Figure 5(d). In addition, when the intensity of the longest ELP loops III starts to decrease, the AR light curve in AIA 335 {\AA} has also fallen back toward the background level. All these observational facts suggest that long-lasting cooling should be the dominant mechanism to produce the ELP in flare F3.

\subsubsection{The September 8 M6.7 and September 9 M2.7 Flares}
Figures 8 and 9 depict the evolution of flares F4 (September 8 M6.7) and F5 (September 9 M2.7), respectively. Like flare F3, both flares are  driven by the eruption of an EW oriented filament FR (pointed out by the arrows in Figures 8(d) and 9(d)) from the similar location under a fan-spine dome. Therefore, similar magnetic reconnections (responsible for the production of either main flare loops or ELP loops) should be involved in the two flares as those in flare F3. Nevertheless, the direction of the FR eruption is different between flares F4 and F5. The FR in flare F4 exhibits a northwestward propagation in the plane of the sky. A large fraction of the cold filament material falls back after the filament reaches its highest altitude (Figure 8(h)), with the runaway part forming a weak CME event with a low projected speed  (214 km s$^{-1}$) and narrow angular width (37$^{\circ}$) in LASCO C2 camera \citep{ZhangQM15}. This may explain why the flare is not accompanied by a type II burst in contrast to other eruptive events. In flare F5, the direction of the FR eruption is instead southeastward, which is further confirmed by the appearance of a corresponding CME in a southeast position angle (254$^{\circ}$) later on. Since the ELP loops are produced by the magnetic reconnection between large-scale overlying field lines stretched out by the erupting FR, the location of the ELP loops should also follow the direction of the FR eruption, which is indeed the case seen in flares F4 and F5. In flare F4,  diffused ELP loops  span diagonally over the northern part of the AR (traced by the dashed line in Figure 8(f)), while in flare F5, the ELP loops are mainly located in the south (indicated by the arrow in Figure 9(f)).

In flares F4 and F5, both the full-disk irradiance in EVE 335 {\AA} (Figures 3(d) and (e)) and the AR-integrated intensity in AIA 335 {\AA} (Figures 8(j) and 9(j)) reveal an elevated post-flare ``background" relative to the pre-flare level, which largely smears out the true contribution from the ELP loops. To extract the  true ELP emissions, we select several sub-regions from the AR (enclosed by the colored boxes in Figures 8 and 9) and track the intensity evolution in AIA 335 {\AA} over these regions (shown in Figures 8(k) and 9(k)), as we have done before.

For flare F4, the intensity of the main flare region (R1 in red) drops from the main phase peak to the pre-flare level finally. Possibly due to a partial eruption of the FR, the region enclosed by the remote ribbon Rr, which typically exhibits a dimming during other eruptive flares, does not show obvious changes in this flare. For this reason, we combine it into the extended ELP loop region (R2 in blue, excluding the extent of region R1). Unlike the flat AR light curve after the main phase peak, the intensity profile of region R2 now reveals a second enhancement of finite duration ($\sim$2 hours), which may constitute the true ELP of the flare. Afterwards, the main contributor to the elevated AR emission appears to switch to a southern region of the AR (R3 in pink). It is seen that region R3 covers a series of long coronal loops emanating from the southwest peripheral of the AR\@. During the flare, the loops continue to brighten up in AIA 335 {\AA} (Figures 8(e) and (f)), while in AIA 171 {\AA}, they show a persistent dimming, which can be easily discerned from the plain images (Figures 8(h) and (i)). The emission pattern indicates a persistent heating of moderate strength in these loops, which mainly raises the loop temperature to a warm coronal temperature (e.g., $\sim3$~MK). Although the loops in region R3 are long relative to the main flare loops, and they brighten up in AIA 335 {\AA} in the late stage, they do not belong to ELP loops like those in region R2. The most important difference is that the loops in region R3 have not been disturbed by the FR eruption, and hence do not participate in any flare-related magnetic reconnections. In passing we note that the heating of these non-ELP loops is more likely due to continuous low atmospheric activities (e.g., magnetic emergence and/or cancellation) taking place at the loop footpoint in AR peripheral, which is certainly irrelevant to the course of the flare.

For flare F5, the situation is quite similar, except that a discernible dimming occurs in the remote dimming region (R3 in purple) as desired. (For this reason here we pick it up separately.) After eliminating the contribution from the remote dimming region and the main flare region (R1 in red), the intensity profile of the ELP loop region (R2 in blue) also reveals a finite duration in ELP enhancement, which lasts for $\sim$2.5 hours. After the ELP peak, the elevated AR emission mainly comes from the brightening of loops in a northern region of the AR (R4 in olive). The position of region R4 is away from the direction of the FR eruption (southwestward in this event). For the same reason, the loops in region R4 are not ELP loops, and the heating mechanism in these non-ELP loops is believed to be identical to that in flare F4.

In both flares, a true ELP peak is revealed from the sub-region light curve. The ELP emission starts to increase shortly after the main phase peak and the time lag between the ELP peak and the main phase peak ($\sim$1 hour) is comparable to the cooling time of a long flaring loop. In AIA images, the ELP loops are observed sequentially from the hot passband to the cool passband. Therefore, the long-lasting cooling mechanism should be responsible for the production of the ELP in these two flares.

\subsubsection{The September 9 M1.2 Flare}
Flare F6 (September 9 M1.2) is the only confined event in our dataset. Evolution of the flare is displayed in Figure 10. Different from all other flares, the flare-driving FR in this flare is initially NS oriented (Figure 10(d)), which implies that the uprising FR would suffer much more confinement imposed by the overlying EW elongated field lines, therefore stable against the torus instability \citep{Kliem06}. This may partially explain why this flare is non-eruptive compared to other eruptive flares.

Based on an NLFFF extrapolation and hydrodynamic modeling, \citet{Dai18b} have proposed a two-stage magnetic reconnection process to  explain the flare evolution. The ascending FR first squashes the above fan-spine structure, and drives reconnections at the null point and the surrounding QSL, producing the remote brightening flare ribbon and long ELP loops (Figure 10(a)). The second-stage reconnection sets in just 5 minutes later; reconnection mainly occurs between the two legs of the field lines stretched by the eventually stopped FR, energizing the compact main flare loops (Figure 10(e)). The two stages of reconnection manifest themselves as two adjacent peaks in the AIA 1600~{\AA} light curve of the AR (solid curve in Figure 10(l)). The first-stage reconnection is believed to be more energetic, therefore resulting in an extremely large ELP (Figures 3(f) and 10(j)).

At the time of flare F6, the AR has rotated to a location close to the western limb. In the main flare region (R1 in red), the AIA 335 {\AA} intensity profile also reveals a large ELP peak (Figure 10(k)), which is mainly contributed by a large amount of ELP loops overlapping along the LOS\@. To reliably trace the evolution of the ELP free of contamination from the main flare region, we select an eastern region (R2 in blue) that only covers the eastern part of the ELP loops. As expected, the intensity in region R2 shows a gradual rise and a fast decay (Figure 10(l)), indicating a ``complete" long-lasting cooling process in  the initially hot ELP loops.

\subsection{Correlation between the Flares}
According to the analysis presented above, the main production mechanism of the ELP for the six homologous flares from AR 11283, changes from additional heating in flare F1 to long-lasting cooling in flares F3--F6, with both mechanisms playing a role in flare F2. As revealed in Table 2, the time lag of the ELP peak in EVE 335 {\AA}, measured with respect to the AIA 1600 {\AA} peak of either the whole flare or the remote ribbon (here we adopt the latter one), shows an abrupt drop after flare F2. The significantly longer time delay of the ELP peak in the first two flares (especially for flare F1) indicates that an additional heating is needed to further postpone the occurrence of the ELP peak.

To further validate the long-lasting cooling scenario for flares F3-F6, we also pick up one representative ELP loop from each event of them (traced by the dashed curves in Figures 7(h) and 8-10(f)), and carry out the same DEM analysis and cooling time estimation of the loops as done in flare F2. The results are listed in Table 3. All these ELP loops experience a brightening-fading-brightening process in AIA 131 {\AA}. The first AIA 131 {\AA} peak appears shortly after the corresponding main flare peak. The DEM analysis at these times (second column) reveals temperatures around 10~MK and densities around $2\times10^{10}$~cm$^{-3}$ for the ELP loops. It is clear that the ELP loops are heated nearly simultaneously with the main flare heating and cool down from an initially high temperature. When they brighten up again in AIA 131 {\AA} (third column), the elapsed times (fourth column) are regarded as the observed loop cooling times $t_{\mathrm{cool}}$. It is found that the observed cooling times of the loops are in good agreement with the theoretically estimated values $\tau_{\mathrm{cool}}$ (sixth column), which are derived by using equation (1), with the relative difference between them being less than 10\%.  This strongly favors the long-lasting cooling process in the ELP loops. In passing we note that according to equation (1), the loop cooling time mainly depends on the loop half-length $L$ (fifth column), whereas it is rather insensitive to the initial temperature and density, as long as the loop is initially hot enough.

We further investigate the correlation between the ELP peak intensity in AIA 335 {\AA}  and the AIA 1600 {\AA} fluence of either the whole flare or the remote ribbon (derived by integrating the solid and dashed curves in Figures 4(o), 6(l), 7(o), and 8(l)--10(l), respectively), a possible relationship analogous to the well-known Neupert effect \citep{Neupert68}. The original Neupert effect relates the time-integrated hard X-ray (HXR) emission of a solar flare to its SXR flux. Using the UV footpoint calorimeter (UFC) method that infers flare heating from AIA 1600 {\AA} light curves at the footpoints of flare loops, recently \citet{Qiu21} has successfully modeled the observed SXR emission in several solar flares. Meanwhile, through 0D hydrodynamic numerical experiments, \citet{Dai18a} found that under a sufficiently strong heating, flare loops will experience a self-similar thermodynamic evolution until they cool down to an intermediate temperature, e.g., 3 MK. In this sense, a correlation between AIA 1600 {\AA} fluence and AIA 335 {\AA} intensity would be of physical significance. The reason for additionally extracting the fluence from the remote ribbon is that most ELP loops are found connecting to the remote ribbon.

To more accurately quantify the ELP emission, we first adopt the method described in \citet{Chen20} to determine the ``exact" spatial extent of the ELP loops. For each flare, we subtract the AIA 335 {\AA} image at the main phase peak from the corresponding image at the ELP peak. In the resultant difference image, pixels with positive values (or more strictly speaking, above a positive threshold value) are included into the ELP region, while those with negative values (below a negative threshold value) belong to the main flare region. The spatial extents of the ELP region (blue) and the main flare region (red) for all six flares are displayed in Figure 11. It is clearly seen that the ELP region occupies a significantly larger area than the main flare region. Based on the figure, the ELP intensity is computed by summing all pixel values over the marked ELP region. To evaluate the uncertainty introduced by the ELP peak time determination, we apply the same operation to all AIA 335 {\AA} images within a 10-minute time window centered at the determined ELP peak time, which results in a total of 51 ELP intensities for each flare. The average of these intensities is taken as the ELP peak intensity, and their standard deviation is regarded as the uncertainty. It is found that the uncertainty of the ELP peak intensity is typically small, within the 3\% level for all flares under study. The reason for the small uncertainty lies in the long lengths of the ELP loops, which make the thermodynamic evolution inside them quite gradual.

The fluence in AIA 1600 {\AA} is in practice calculated by integrating the intensity over the impulsive rising phase of the flare light curve, since the decay phase may be somewhat affected by non-heating effects. We set a threshold of 10 times the background fluctuation to define the start point of the impulsive phase. Since here the background level of all AIA 1600 {\AA} light curves is very steady, using different thresholds in a wide range actually does not affect the final results. In addition, due to the complexity of the flares, here the whole flare light curves typically exhibit multiple spikes around the peak. Therefore, in deriving the whole flare fluence, we alternatively integrate the AIA 1600 {\AA} intensity until it drops to half the peak.

Scatter plots of the quantities calculated above are shown in Figure 12. In both plots, with the aid of a power-law fitting (red line), a positive correlation between the ELP peak intensity and the flare ribbon fluence is established for the five  eruptive flare events. The correlation coefficient for the case of the remote ribbon fluence (0.946, lower panel) is considerably higher than that for the case of the whole flare fluence (0.811, upper panel), indicating a much closer physical link between the brightening of the remote ribbon and the heating of the ELP loops as expected. Since the remote ribbon attains the peak nearly simultaneously with other flare ribbons, this high correlation coefficient strongly favors the long-lasting cooling scenario in producing the ELP\@. At first glance, this argument seems to contradict the ELP production scenario in flare F1. We note that both the ELP peak intensity and the remote ribbon fluence in flare F1 are the weakest  among all flares in our dataset. Hence the proximity of the data point for flare F1 to the fitting line should just be a coincidence. It is also worth noting that in the lower panel, the data point for flare F2 lies well above the fitting line. It is not hard to understand, since the additional heating in this flare may further enhance the ELP peak.

As to the only non-eruptive flare F6, the corresponding data point is totally an outlier in the upper panel, but moves closer to the fitting line in the lower panel. It basically indicates a long-lasting cooling process taking effect in flare F6. However, for this confined flare, the significantly higher ELP peak intensity compared to that inferred from the correlation may imply a different energy partitioning inside the ELP loops.

\section{Discussion and Conclusions}
Using the observations with \emph{SDO}, we investigate the production of the ELP in six homologous flares (F1--F6) originating from a complex AR NOAA 11283, with an emphasis on the emission characteristics of the flares. It is found that the main production mechanism of the ELP changes from additional heating in flare F1 to long-lasting cooling in flares F3--F6, with both mechanisms playing a role in flare F2. This transition is evidenced by an abrupt decrease of the time lag of the ELP peak. Furthermore, a positive correlation between the flare ribbon fluence and the ELP peak intensity is established in the majority of the flares, suggesting a long-lasting cooling process following an early heating in the ELP loops.

The change in ELP production mechanism after flare F1 may be attributed to an enhancement of the envelope (or overlying) magnetic field above the AR, which not only brings forward the time of the interaction between the rising FR and the large-scale overlying magnetic field lines, but also enhances the strength of the interaction. Compared to the heating in flare F1, the enhanced interaction in the following flares causes a more prompt and energetic heating of the ELP loops, leading to a more notable ELP after a sufficiently long cooling time. Evidence of the enhancement of envelope magnetic field relies on a detailed investigation of the AR magnetic evolution, which is beyond the scope of this work, and we will present it in a separate work.

The orientation of the ascending FR is another important factor affecting the interaction between the FR and the overlying field lines. When the FR is mainly NS oriented, it will either push or stretch more overlying EW elongated field lines, consequently more effectively energizing the ELP loops. It is noted that an NS oriented FR was involved in flares F3 (after a clockwise rotation) and F6. As revealed in Figure 12, these two flares produce the largest ELP peaks among all flares. In addition, the enhanced interaction also imposes a higher confinement on the rising FR. If the FR is not energetic enough, its ascent would be more subject to the torus instability \citep{Kliem06}, resulting in a failed eruption. This may partially explain why the least energetic (according to the \emph{GOES} classification) flare F6 in our dataset is non-eruptive compared to other eruptive flares.

In our dataset, the only confined flare F6 produces an extremely large ELP peak (relative to the main flare peak). This is consistent with the argument in \citet{WangYM16} that a strong constraint of the overlying arcades may prevent an FR from escaping, and cause the energy carried by the FR to be re-deposited into the thermal emissions that form a stronger late phase. Interestingly, the flare heating does not power compatibly strong brightening in the chromospheric ribbon (see Figure 12). We attribute it to a different energy partitioning in the ELP loops for this confined flare. Different from other eruptive flares, the ELP loops in this flare are mainly heated by slipping or slip-running reconnection taking place in the QSL \citep{Dai18b}. In such QSL reconnection, the released magnetic energy should be more likely to be converted to thermal component through ohmic dissipation rather than to non-thermal electrons. Compared to thermal component, the non-thermal electron bombardment can cause more significant responses in chromospheric radiation \citep[c.f.][]{Hong19}, although the two components may have a similar effect in raising the coronal emission.

\begin{acknowledgments}
We are very grateful of the anonymous referee for constructive comments and suggestions that led to a significant improvement of the manuscript. This work was supported by National Natural Science Foundation of China under grants 11533005 and 11733003. Y.D. is also sponsored by National Key R\&D Program of China under grants 2019YFA0706601 and 2020YFC2201201. \emph{SDO} is a mission of NASA's Living With a Star (LWS) Program.
\end{acknowledgments}


\clearpage
\begin{deluxetable}{ccccccccc}
\caption{List of the Flares under Study and Their Associated Phenomena}
\label{tab-table}
\tablehead{
\multicolumn{6}{c}{Flare} & \colhead{} & \multicolumn{2}{c}{Other Phenomena}\\
\cline{1-6}  \cline{8-9}
\colhead{No.} & \colhead{Date} & \colhead{Start} & \colhead{Peak} & \colhead{Class} & \colhead{Location} & \colhead{} & \colhead{CME} & \colhead{Type II Burst}
}
\startdata
F1  & Sep-06  & 01:35  & 01:50  & M5.3  & N14W07  &  & Yes  & Yes \\
F2  & Sep-06  & 22:12  & 22:20  & X2.1  & N14W18  &   & Yes  & Yes \\
F3  & Sep-07  & 22:32  & 22:38  & X1.8  & N14W28  &   & Yes  & Yes \\
F4  & Sep-08  & 15:32  & 15:46  & M6.7  & N14W40  &   & Yes  & No \\
F5  & Sep-09  & 06:01  & 06:11  & M2.7  & N14W47  &   & Yes  & Yes  \\
F6  & Sep-09  & 12:39  & 12:49  & M1.2  & N13W52  &   & No  & No
\enddata
\end{deluxetable}

\clearpage
\begin{deluxetable}{ccccc}
\caption{Peak Time Information of the Flares}
\label{table2}
\tablehead{
    \colhead{No.} & \colhead{Whole Flare Peak} & \colhead{Remote Ribbon Peak}  &  \colhead{ELP Peak} & \colhead{Time Lag\tablenotemark{a}} \\
    \colhead{} & \colhead{\footnotesize AIA 1600 {\AA} (UT)} & \colhead{\footnotesize AIA 1600 {\AA} (UT)} & \colhead{\footnotesize EVE 335 {\AA}  (UT)} & \colhead{(min)}
}
\startdata
F1  & Sep-06 01:44:41 & Sep-06  01:44:17  & Sep-06 03:58:24  & 134  \\
F2  & Sep-06 22:18:41 & Sep-06  22:19:53  & Sep-06 23:49:24  & 89.5 \\
F3  & Sep-07 22:36:17 & Sep-07  22:37:05  & Sep-07 23:47:44  & 70.7 \\
F4  & Sep-08 15:40:41 & Sep-08  15:40:17  & Sep-08 16:40:04\tablenotemark{b} & 59.8\\
F5  & Sep-09 06:07:53 & Sep-09  06:05:53  & Sep-09 07:18:40\tablenotemark{b} & 72.8\\
F6  & Sep-09 12:42:41 & Sep-09  12:42:41  & Sep-09 13:38:04  & 55.4
\enddata
\tablenotetext{a}{\footnotesize Time delay of the ELP peak with respect to the peak of the remote ribbon intensity.}
\tablenotetext{b}{\footnotesize Peak time alternatively determined from the intensity of the ELP region in AIA 335 {\AA}.}
\end{deluxetable}

\clearpage
\begin{deluxetable}{ccccccc}
\caption{Observed and Theoretically Estimated  Cooling Time of the Representative ELP Loops in Flares F3--F6}
\label{table3}
\tablehead{
    \colhead{Parent Flare} & \colhead{ AIA 131 {\AA} (hot)} &  \colhead{ AIA 131 {\AA} (cool)} & \colhead{$t_{\mathrm{cool}}$\tablenotemark{a}} & \colhead{$L$} & \colhead{$\tau_{\mathrm{cool}}$\tablenotemark{b}} \\
    \colhead{} & \colhead{\footnotesize (UT)} & \colhead{\footnotesize  (UT)} & \colhead{(min)} & \colhead{(Mm)} & \colhead{(min)}
}
\startdata
F3  & Sep-07 22:39:59 & Sep-08 00:17:57  & 97.0 & 88.8 &  96.9 \\
F4  & Sep-08 16:12:35 & Sep-08 17:59:01  & 106 & 92.8 & 109 \\
F5  & Sep-09 06:17:49 & Sep-09 07:43:21  & 85.5 & 61.5 & 77.4 \\
F6  & Sep-09 13:03:11 & Sep-09 14:01:33  & 58.4 & 51.1 & 58.7
\enddata
\tablenotetext{a}{\footnotesize Observationally determined cooling time from the AIA 131 {\AA} peaks.}
\tablenotetext{b}{\footnotesize Theoretically estimated cooling time with equation (1).}
\end{deluxetable}

\clearpage
\begin{figure}
\epsscale{1}
\plotone{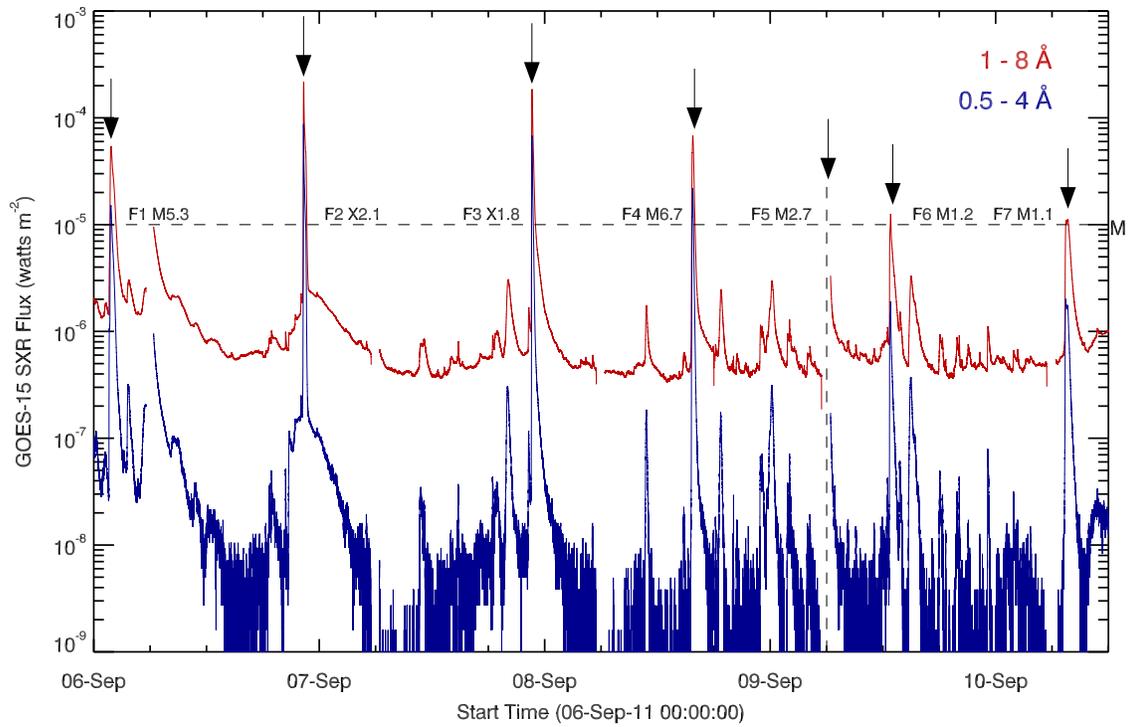}
\caption{Time profiles of the \emph{GOES} fluxes in the long (1--8 \AA, red) and short bands (0.5--4 \AA, blue), respectively. The arrows point out 7 M-and-above flares (F1--F7) originating from AR 11283, with the class of each flare labeled beside the flare peak. Note that the peak of flare F5 is missed by \emph{GOES} due to orbit shading, so the peak time (vertical dashed line) and peak intensity of the flare are inferred from the observations with other instruments that cover the similar wavelength range.}
\end{figure}

\clearpage
\begin{figure}
\epsscale{1}
\plotone{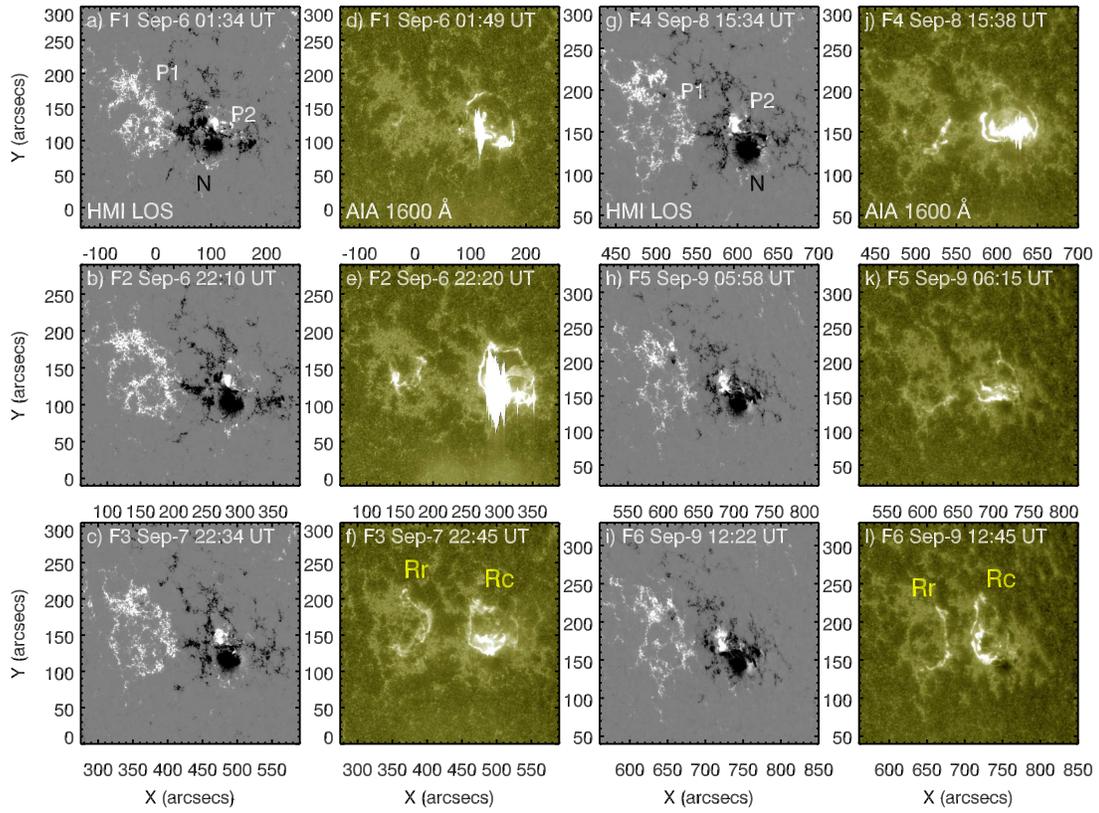}
\caption{HMI LOS magnetogram of AR 11283 before each flare (a--c and g--i), and AIA 1600 {\AA} image of the corresponding flare around the flare peak (d--f and j--l). P1, P2, and N represent three main polarities of the AR, while Rc (Rr) denotes a circular (remote) flare ribbon located in polarity N (P1).}
\end{figure}

\clearpage
\begin{figure}
\epsscale{1}
\plotone{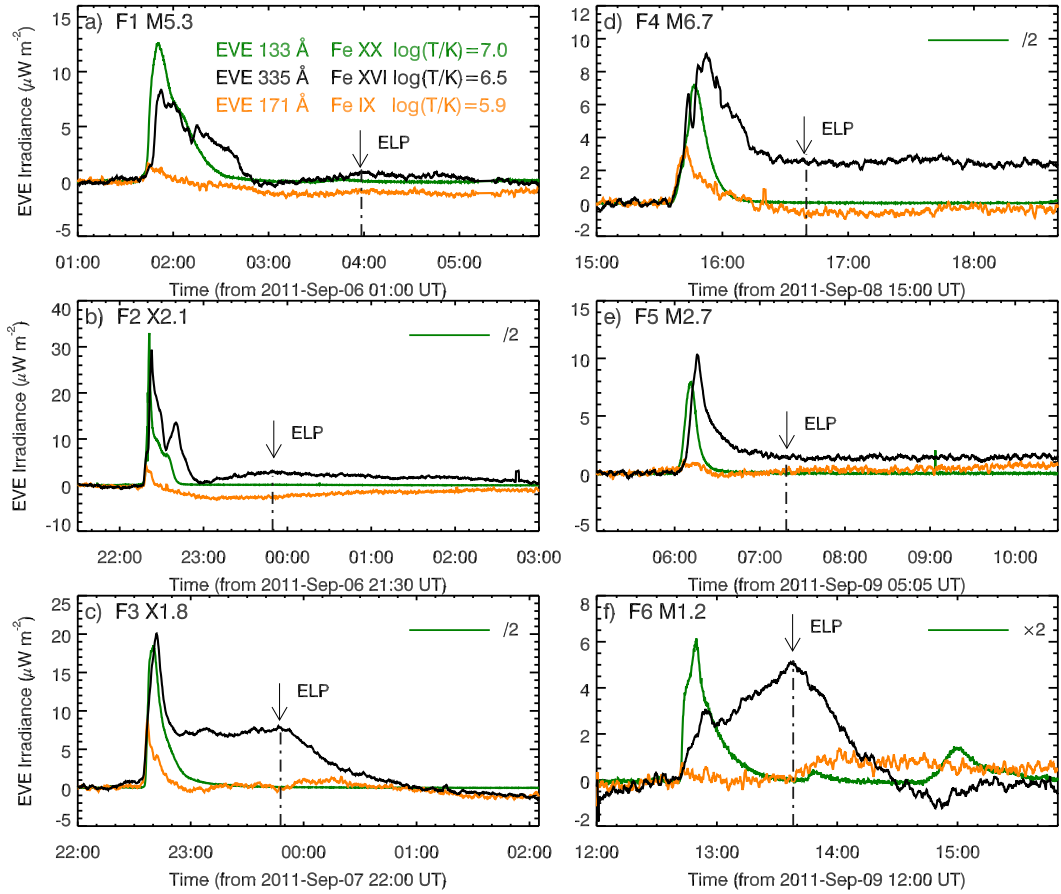}
\caption{Background-subtracted irradiance of the flares under study in three EVE lines labeled in the legend. The arrows and dash-dot lines outline the ELP peak of the \ion{Fe}{16} 335 {\AA} line irradiance (black) for each flare. Note that the ELP peaks in flares F4 and F5 are inferred from the spatially resolved AIA observations.}
\end{figure}

\clearpage
\begin{figure}
\epsscale{1}
\plotone{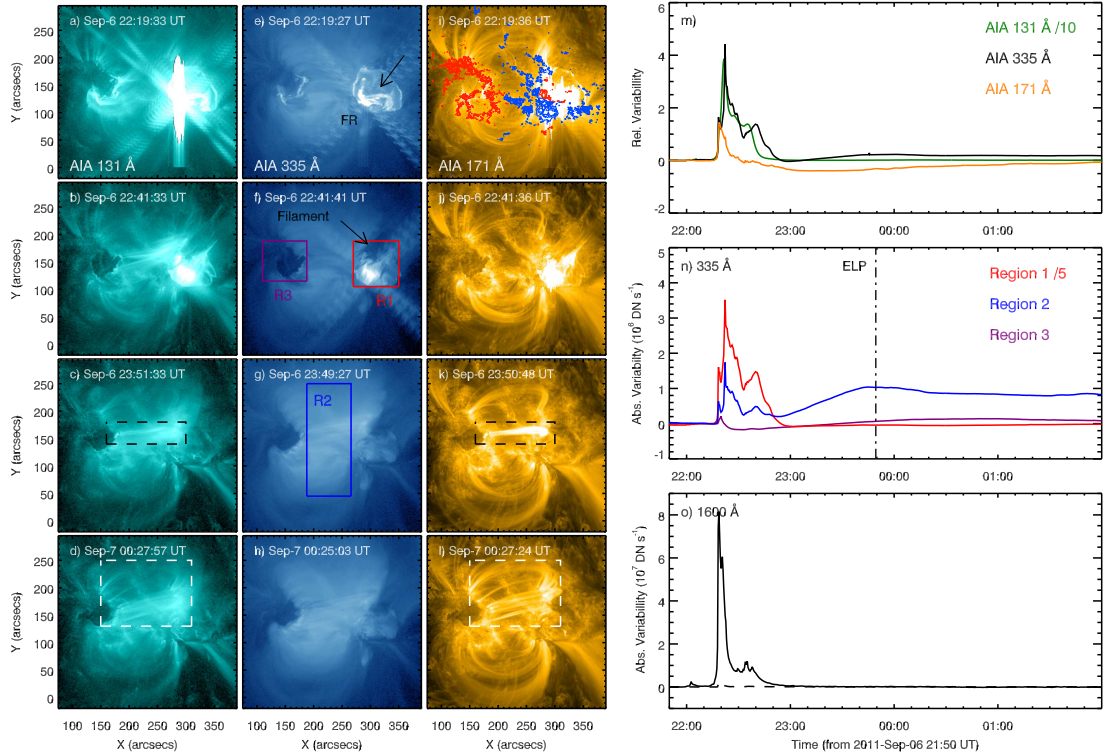}
\caption{Evolution of the September 6 X2.1 flare. Left: Images of the flare taken by AIA in 131 (a)--(d), 335 (e)--(h), and 171 {\AA} (i)--(l), respectively. Over-plotted on panel (i) are the contours of a pre-flare HMI LOS magnetogram at levels of $\pm$250, $\pm$600, and $\pm$950 G (red for positive, and blue for negative). Some characteristic structures during the flare evolution are highlighted by the arrows or dashed rectangles; see the text for a detailed description. Right: Light curves of the flare in different AIA passbands. Panel~(m) displays the AR relative variabilities (with the background-subtracted value scaled by the corresponding background level) in AIA 131, 335, and 171 {\AA}, respectively. Panel (n) shows the background-subtracted AIA 335~{\AA} intensity profiles in three sub-regions (labeled as R1, R2, and R3, respectively) outlined by the colored boxes in panels (f) and (g). Panel (o) displays the background-subtracted AIA 1600 {\AA} intensity profiles for the whole AR (solid) and the remote flare ribbon region (dashed). The vertical dash-dot line in panel~(n) denotes the ELP peak of the full-disk irradiance in EVE 335 {\AA} (23:49 UT), which is nearly simultaneous with the ELP peak of the AR intensity in AIA 335 {\AA}, and is also very close to the ELP peak of the intensity in region R2.}
\end{figure}

\clearpage
\begin{figure}
\epsscale{1}
\plotone{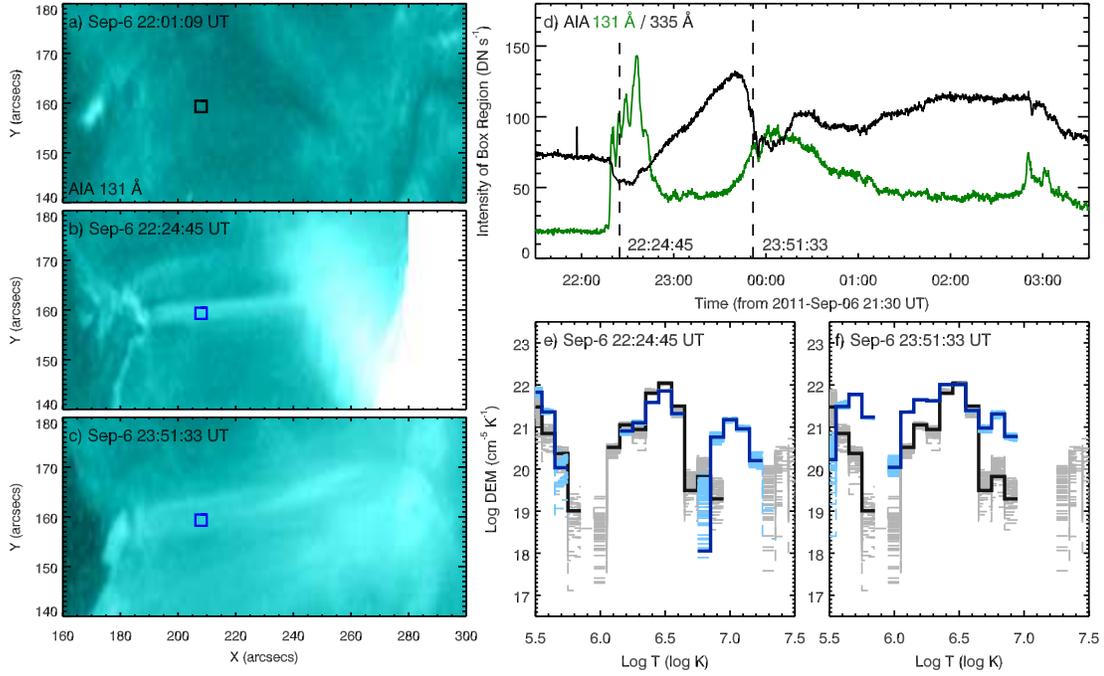}
\caption{Left: AIA 131 {\AA} images taken at the pre-flare (a), the early (b), and the late (c) stage of the flare evolution in the September 6 X2.1 flare, showing the heating and cooling process of a representative ELP loop. The field-of-view (FOV) of the panels is outlined by the black dashed rectangle in Figure 4(c). A small box is drawn on the loop spine. Upper right: Intensity profiles of the selected box region in AIA 131 and 335 {\AA}. The vertical dashed lines indicate two local peaks in the AIA 131 {\AA} light curve, which correspond the observation times of panels (b) and (c). Lower right: DEM distributions of the box region at the two AIA 131 {\AA} peak times (blue solid histogram) as well as at the pre-flare time (black solid histogram). In each panel, the over-plotted hatched histograms are the results of 100 Monte Carlo runs by randomly adjusting the inputs within the corresponding standard deviation.}
\end{figure}

\clearpage
\begin{figure}
\epsscale{1}
\plotone{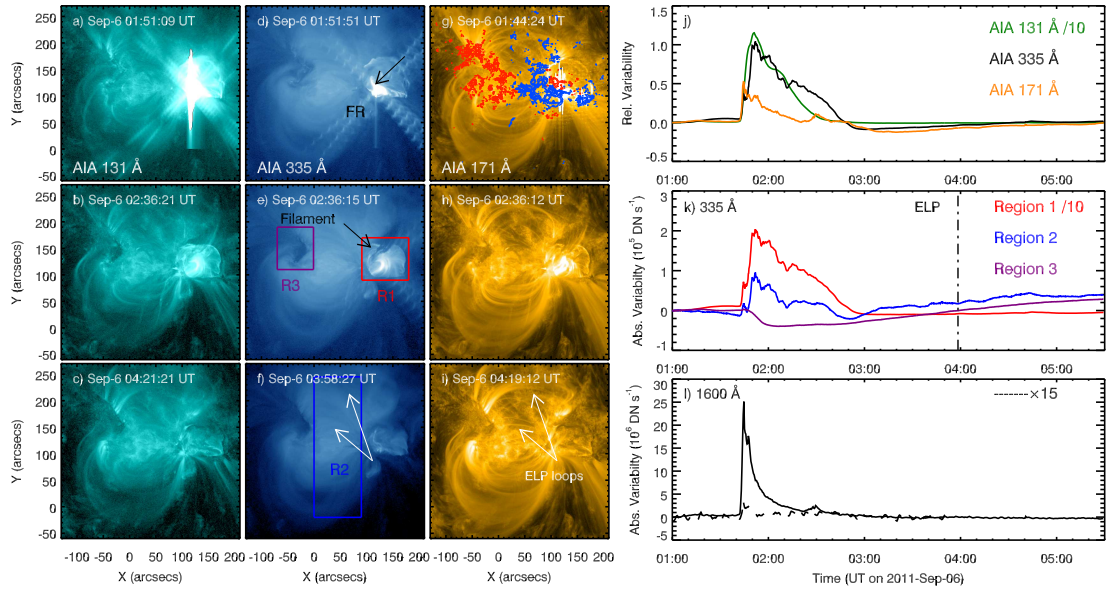}
\caption{Evolution of the September 6 M5.3 flare.  The panels are organized the same as in Figure 4.}
\end{figure}

\clearpage
\begin{figure}
\epsscale{1}
\plotone{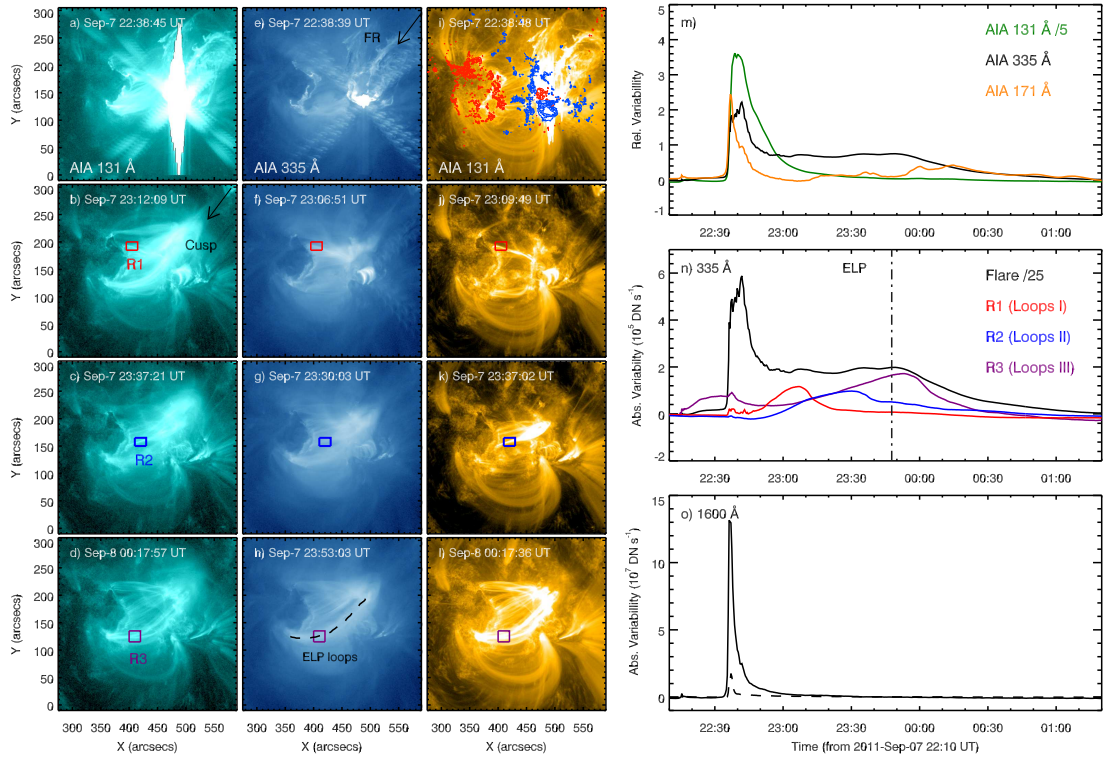}
\caption{Evolution of the September 7 X1.8 flare. The panels are organized the same as in Figure 4. Note that the dashed curve in panel (h) traces a representative ELP loop, and the background-subtracted AR light curve in AIA 335 {\AA} (black) is also over-plotted in panel (n).}
\end{figure}

\clearpage
\begin{figure}
\epsscale{1}
\plotone{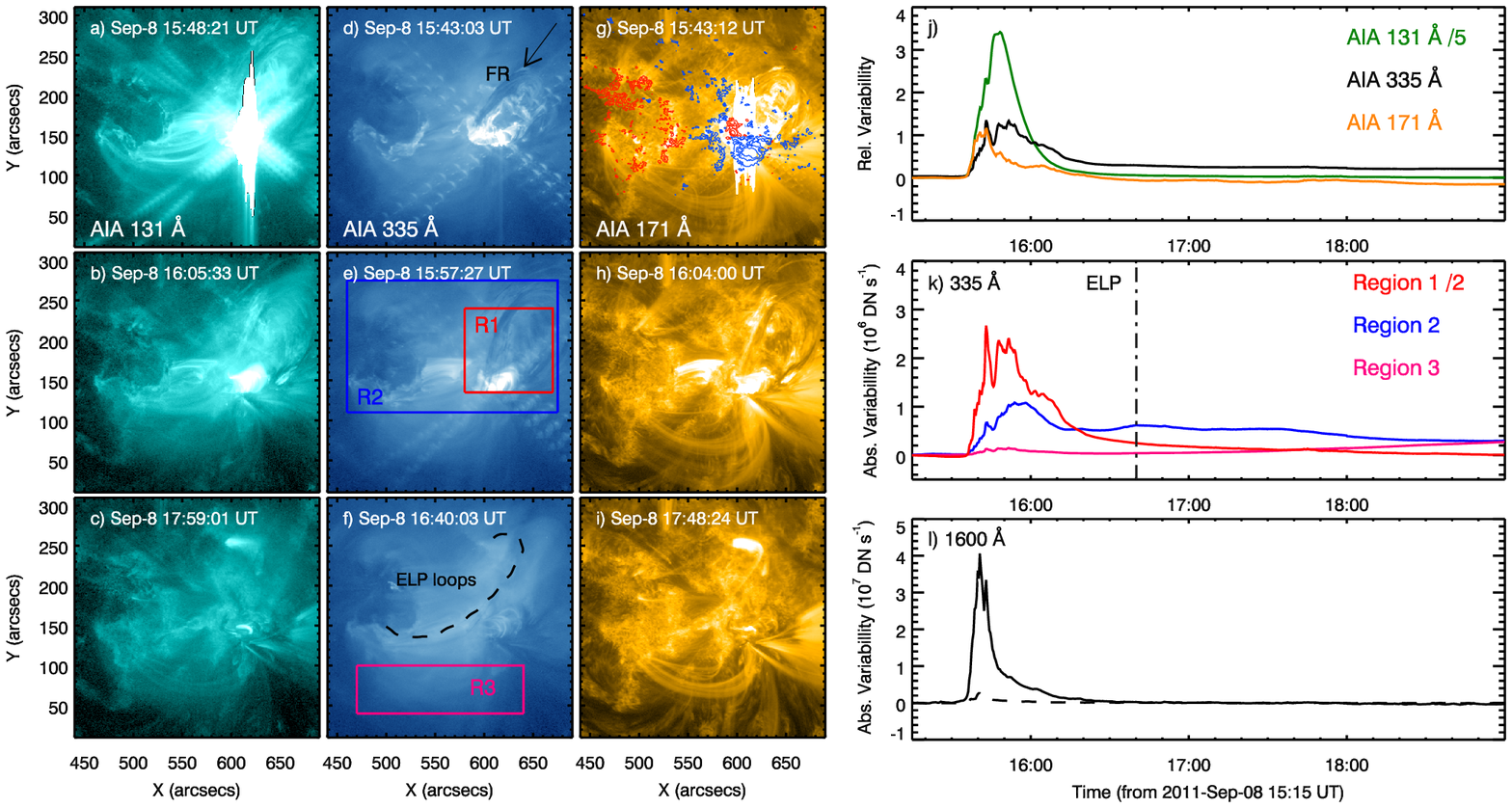}
\caption{Evolution of the September 8 M6.7 flare. The panels are organized the same as in Figure 4 except that the vertical dash-dot line in panel~(k) indicates the ELP peak of region R2 in AIA 335 {\AA}. Note that the dashed curve in panel (f) traces a representative ELP loop.}
\end{figure}

\clearpage
\begin{figure}
\epsscale{1}
\plotone{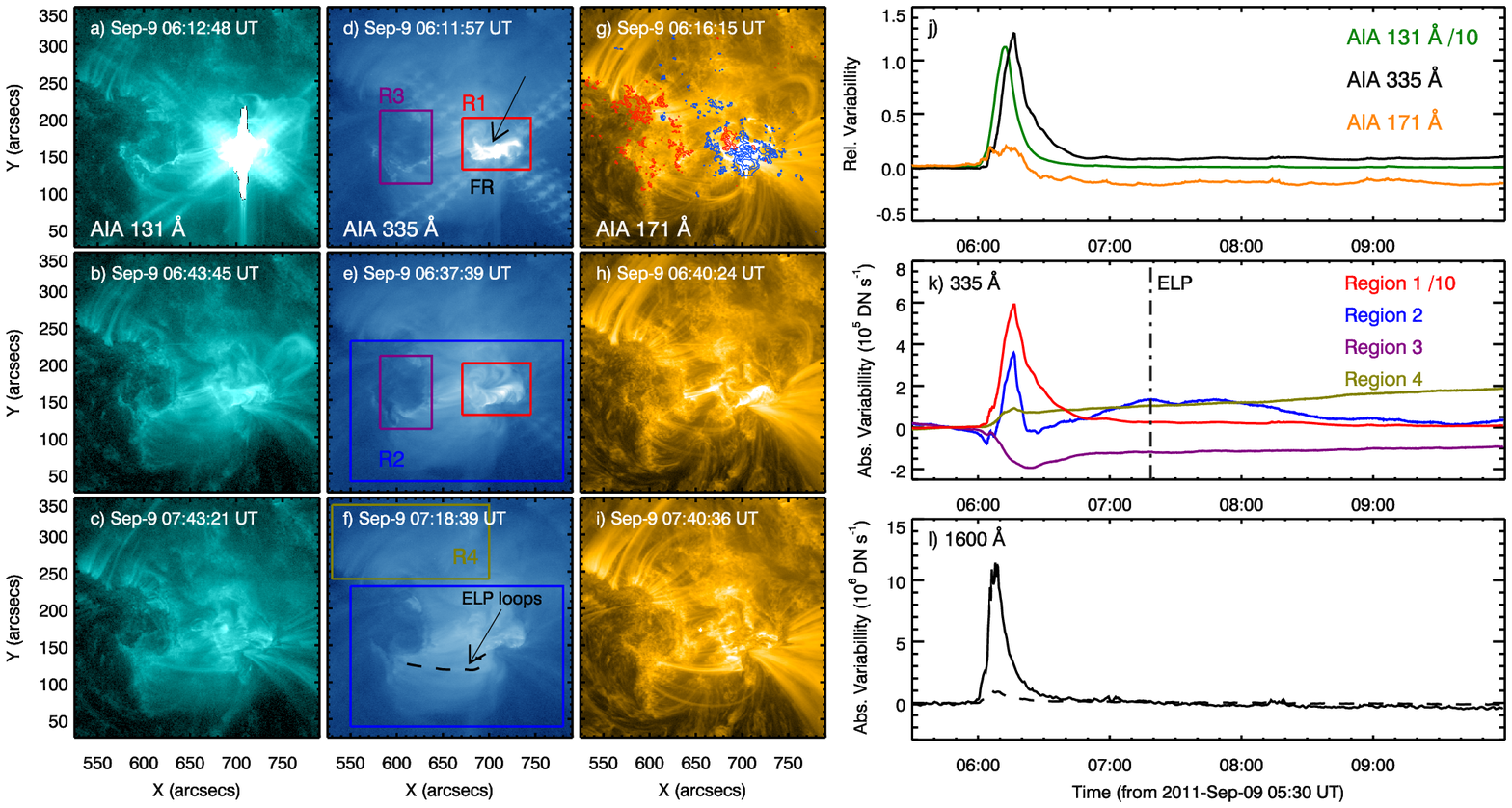}
\caption{Evolution of the September 9 M2.7 flare. The panels are organized the same as in Figure 4 except that the vertical dash-dot line in panel~(k) indicates the ELP peak of region R2 in AIA 335 {\AA}. Note that the dashed curve in panel (f) traces a representative ELP loop.}
\end{figure}

\clearpage
\begin{figure}
\epsscale{1}
\plotone{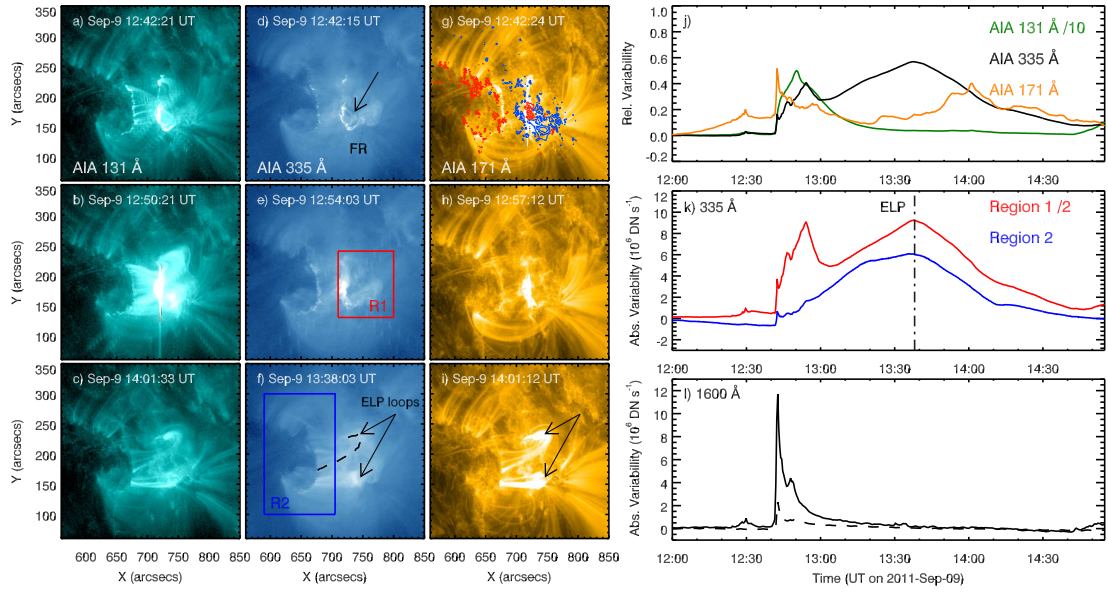}
\caption{Evolution of the September 9 M1.2 flare. The panels are organized the same as in Figure 4. Note that the dashed curve in panel (f) traces a representative ELP loop.}
\end{figure}

\clearpage
\begin{figure}
\epsscale{1}
\plotone{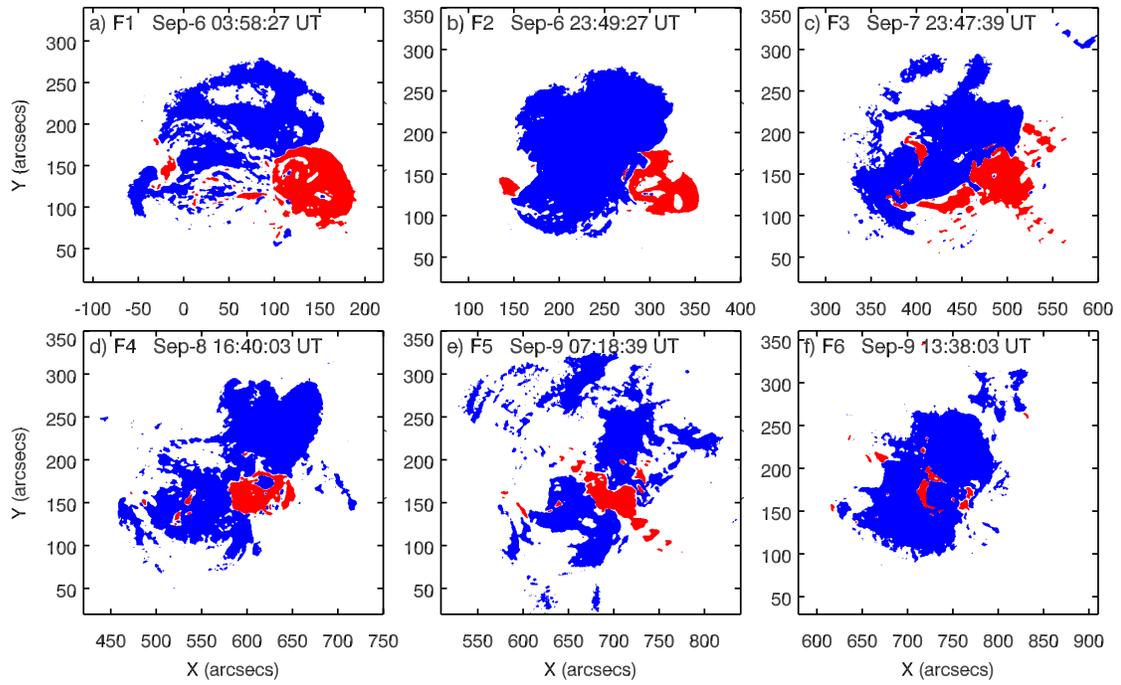}
\caption{Spatial extents of the ELP region (blue) and the main flare region (red) in the six flares.}
\end{figure}

\clearpage
\begin{figure}
\epsscale{1}
\plotone{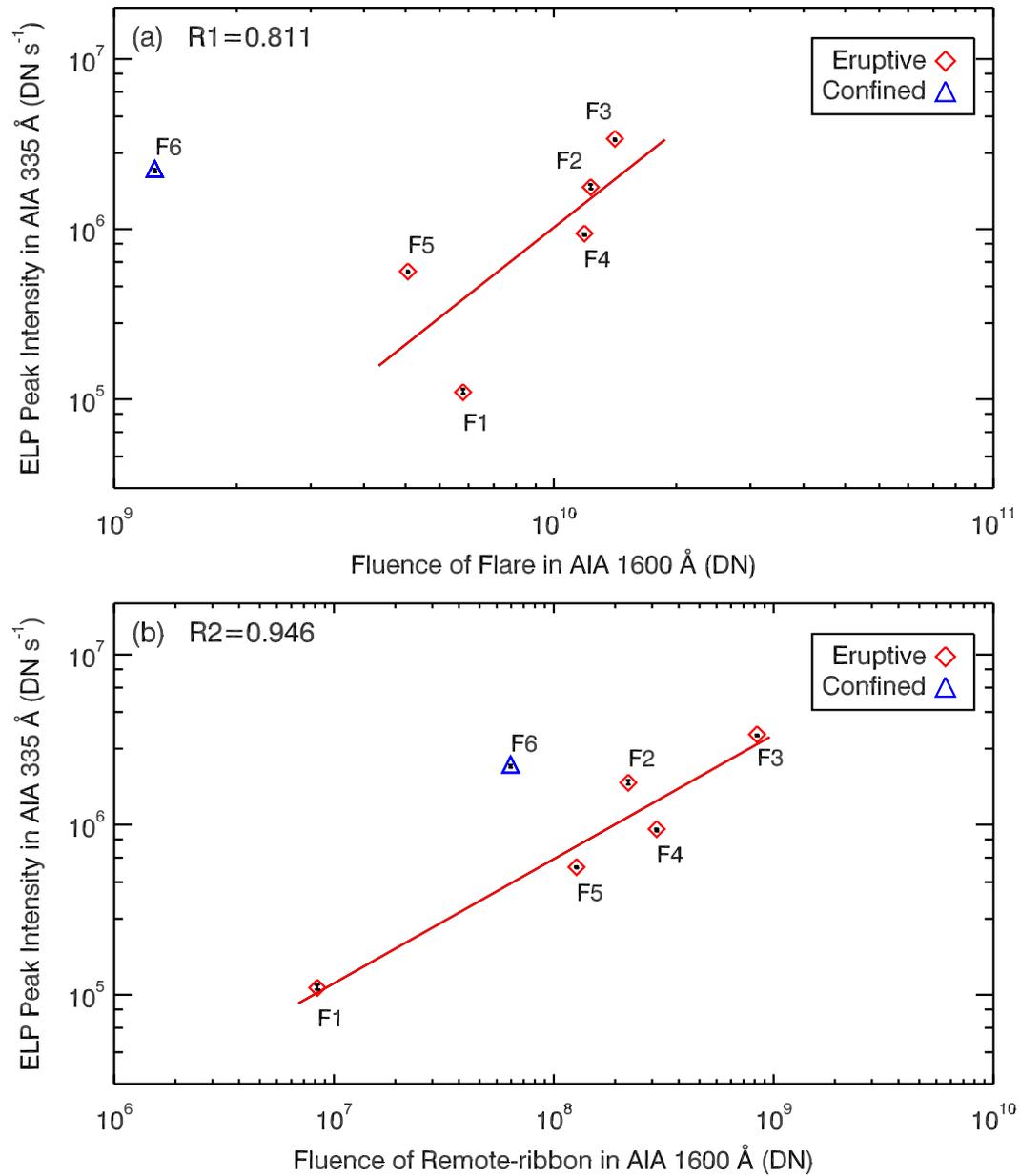}
\caption{Scatter plot of the ELP peak intensity in AIA 335 {\AA} against the AIA 1600 {\AA} fluence of either the whole flare (upper) or the remote ribbon (lower). The red lines denote the best power-law fitting to the data points for the five eruptive events, while $R_1$ and $R_2$ are the corresponding Pearson correlation coefficients. Note that the vertical error bars over-plotted on the data points indicate the uncertainty in ELP peak intensity.}
\end{figure}

\listofchanges

\end{document}